\documentclass[aps,manuscript]{revtex4}
\usepackage[latin1]{inputenc}
\usepackage{amsmath}

\begin{document}

\title{Statistical entropy of three-dimensional q-deformed Kerr-de Sitter
space}

\author{Michael C. Abbott and David A. Lowe}

\email{michael_abbott@brown.edu, lowe@brown.edu}

\affiliation{Physics Department, Brown University, Providence, RI 02912, USA}

\begin{abstract}
A quantum deformation of three-dimensional de Sitter space was proposed
in hep-th/0407188. We use this to calculate the entropy of Kerr-de
Sitter space, using a canonical ensemble, and find agreement with
the semiclassical result.
\end{abstract}
\maketitle

\section{introduction}

Black holes are known to carry entropy proportional to their horizon
area \cite{Bekenstein:1973ur,Hawking:1974sw}. One of the main successes
of string theory has been to provide a microscopic interpretation
of this entropy \cite{Strominger:1996sh}, at least in certain cases.
Since the semiclassical arguments for this horizon area are not specific
to horizons surrounding black holes, they should also apply to cosmological
horizons \cite{Figari:1975km,Unruh:1976db,Gibbons:1977mu}. It would
be of great interest to have microscopic state-counting arguments
for such situations. 

The natural place to investigate cosmological horizons is de Sitter
space, the maximally symmetric spacetime with constant positive cosmological
constant $\Lambda$ \cite{Parikh:2002py,Strominger:2001pn,Banks:2003cg,Banks:2006rx,Banks:2002wr}.
Current observations suggest that our universe is now $\Lambda$-dominated
, and thus asymptotically de Sitter in the future \cite{wmap3-cosmology,Carroll:2000fy,Fischler:2001yj}.

Following the great success of the AdS/CFT correspondence \cite{Maldacena:1997re},
there have been suggestions of a dual conformal field theory, living
on the conformal boundary of de Sitter \cite{Park:1998qk,Strominger:2001pn}.
Unlike the anti-de Sitter case, this boundary is spacelike, and time-translation
in the bulk corresponding to scale transformation on the boundary
\cite{Balasubramanian:1999jd,deBoer:1999xf,Strominger:2001gp}. It
also has two disconnected components, and it is not clear whether
the boundary theory should live on one or both \cite{Parikh:2002py}. 

Since the area of an observer's cosmological horizon is finite, de
Sitter has finite Bekenstein-Hawking entropy. This immediately causes
problems with finding a state counting interpretation, since the isometry
group is non-compact \cite{Goheer:2002vf} and hence only has infinite
dimensional unitary representations. This apparent contradiction is
even stronger if the dimension of the Hilbert space is also finite
\cite{Banks:2000fe,Balasubramanian:2001rb,Banks:2005bm}. It has been
suggested that the correct inner product is not the naive local one,
which changes the notion of unitarity \cite{Witten:2001kn,Bousso:2001mw}.
(Another approach is given in \cite{Parikh:2004ux,Parikh:2004wh}.) 

Some of these difficulties might be tamed by noncommutative geometry
\cite{connes94,Madore:1999bi}. The approach used in \cite{Guijosa:2003ze}
is to deform the group of isometries into a quantum group. This was
further studied in \cite{Lowe:2004nw,Guijosa:2005qi,Lowe:2005de}
and is the approach followed here. Quantum deformations of the Lorentz
group in various dimensions are studied in \cite{Woronowicz-94,Buffenoir:1997ih}
and in \cite{Krishnan:2005ct,Krishnan:2006bq}.

The plan of the paper is as follows. Section \ref{sec:q-Deformed-de-Sitter}
of this paper is mostly a recap of \cite{Guijosa:2003ze,Lowe:2004nw};
section \ref{sec:Entropy-of-Kerr--de} contains a novel calculation
of the entropy. The interpretation of the result is discussed and
concluding remarks made in section \ref{sec:Conclusions}.

\section{q-Deformed de Sitter space\label{sec:q-Deformed-de-Sitter}}

Three-dimensional de Sitter space can be defined as the hyperboloid
$-(x^{0})^{2}+\sum_{i=1}^{3}(x^{i})^{2}=\ell^{2}$ in Minkowski space.
This is a spacetime of constant curvature, representing a vacuum with
positive cosmological constant $\Lambda=1/\ell^{2}$. 

The isometries of this hyperboloid are just the rotations and boosts
of the embedding space, which generate the Lorentz group $SO(3,1)$.
We will focus on the Lie algebra, rather than the global properties
of the group. The complex combinations of generators $X_{i}=J_{i}+iK_{i}$
(left) and $\overline{X}_{i}=J_{i}-iK_{i}$ (right) each obey the
$su(2)$ commutation relation $[X_{i},X_{j}]=i\epsilon_{ijk}X_{k}$,
and commute with each other. 

In the complex algebra, the fact that $J$ and $K$ are Hermitian
is encoded in the star operation $J^{\star}=J$, $K^{\star}=K$, and
so $X_{i}^{\star}=\overline{X}_{i}$. The use of this star-structure
specifies that we are dealing with the non-compact real form $so(3,1)$.
We will also use the basis given by: \begin{align*}
L_{0} & =X_{1} & \overline{L}_{0} & =-\overline{X}_{1}\\
L_{1} & =X_{2}-iX_{3} & \overline{L}_{1} & =-\overline{X}_{2}-i\overline{X}_{3}\\
L_{-1} & =-X_{2}-iX_{3} & \overline{L}_{-1} & =\overline{X}_{2}-i\overline{X}_{3}.\end{align*}
These generators form the $n=0\pm1$ part of the Virasoro algebra
$[L_{m},L_{n}]=(m-n)L_{m+n}$, but with real form \begin{equation}
L_{n}^{\star}=-\overline{L}_{n}.\label{eq:so31}\end{equation}

A field in de Sitter space will transform under isometries in some
representation of this algebra. Since the group is non-compact, there
are no finite-dimensional unitary representations, thus any field
has infinitely many modes. For a field of mass $m>\ell$ the representation
is in the principal series \cite{naimark64,vilenkin91,Joung:2006gj}.

It was proposed in \cite{Guijosa:2003ze} that the Lie algebra of
isometries should be deformed to a quantum group (Hopf algebra) \cite{Biedenharn:1996vv,Klimyk:1997eb}.
Taking the deformation parameter to be a root of unity \[
q=e^{2\pi i/N}\]
limits the dimension of an irreducible representation to at most $N$.
In particular, the deformed versions of non-compact algebras can have
finite dimensional unitary representations, which become infinite
in the classical limit $q\to1$ \cite{Dobrev:1993gi,Steinacker:1999fj}.
This was done explicitly for dS$_{2}$'s $so(2,1)$ principal series
in \cite{Guijosa:2003ze}. The relation between $N$ and gravity quantities
will be fixed momentarily.

In dS$_{3}$ however there is a complication which does not arise
in dS$_{2}$: even the deformed algebra cannot have non-trivial unitary
representations \cite{Lowe:2004nw}. Suppose $\left|\psi\right\rangle $
is an eigenstate of $L_{0}$ and $\overline{L}_{0}$ in a unitary
representation. Then the state $L_{\pm1}\left|\psi\right\rangle $
has zero norm, since $L_{1}^{\star}$ does not lower the eigenvalue
$L_{1}$ raised. So the representation must be trivial. (In the infinite-dimensional
principal series representation, such a $\left|\psi\right\rangle $
lies outside the Hilbert space.) 

Similar problems with unitarity are found in \cite{Krishnan:2006bq}
in attempting to deform this and higher Lorentz groups, and multi-parameter
families of deformations were studied in \cite{Krishnan:2005ct}.

These algebraic problems are related to the problem of defining an
inner product for fields on de Sitter space, which in turn induces
a particular adjoint. The standard local Klein-Gordon one, which induces
\eqref{eq:so31}. Witten proposed to use the path integral from asymptotic
past to future, with an extra insertion of $CPT$ \cite{Witten:2001kn}.
Choosing the parity operation to be  $P\, x^{3}=-x^{3}$, \cite{Lowe:2004nw}
showed that this induces  \begin{equation}
L_{n}^{\dagger}=-L_{n},\qquad\overline{L}_{n}^{\dagger}=-\overline{L}_{n}\label{eq:split-form}\end{equation}
or $X_{1,2}^{\dagger}=-X_{1,2}$, $X_{3}^{\dagger}=X_{3}$ and the
same on the right. This amounts to using the the (non-compact) split
real form $su(1,1)\oplus su(1,1)$, instead of $so(3,1)$.%
\footnote{Throughout this paper we use $\star$ for the so(3,1) involution \eqref{eq:so31},
and $\dagger$ for this one. %
}

With this real form, the natural deformation of the algebra to use
is \[
U_{q}\left(su(1,1)\right)\oplus U_{q}\left(su(1,1)\right).\]
The quantum group $U_{q}\left(su(1,1)\right)$ has unitary representations
of dimension $N$. These are representations without highest weight,
having $(X_{\pm})^{N}\neq0$, and are called cyclic representations
($\mathcal{B}$ in \cite{Biedenharn:1996vv}). It was shown in \cite{Guijosa:2003ze}
that the parameters of a cyclic representation can be chosen so as
to give the same Casimirs as the classical $su(1,1)=so(2,1)$ principal
series, and in \cite{Lowe:2004nw} that a left-right product of two
cyclic representations has the correct Casimirs to match the $so(3,1)$
principal series. 

The geodesics lying in the embedding space's 0-1 plane are the north
and south poles of de Sitter space. The south pole is $r=0$ in  the
static coordinate patch, whose metric is\begin{equation}
ds^{2}=-\left(1-\frac{r^{2}}{\ell^{2}}\right)dt^{2}+\frac{dr^{2}}{1-r^{2}/\ell^{2}}+r^{2}d\phi^{2}.\label{eq:static-metric}\end{equation}
The generator of time translations here is\[
-i\partial_{t}=K_{1}=-i\left(L_{0}+\overline{L}_{0}\right).\]
At the antipodal point $-x^{\mu}$ this generates instead reverse
time translation. (This is the standard situation for a thermofield
double, the canonical example of which is Rindler space.) 

In these coordinates the horizon is at $r=\ell$. It has Hawking temperature
$T=1/2\pi$ which can be derived most transparently for our purposes
by tracing over modes living behind the horizon (which have negative
frequency) to produce southern density matrix \cite{Bousso:2001mw}
\[
\rho^{\mathrm{south}}\propto e^{-\beta K_{1}}.\]

In the classical (principal series) case this operator has a continuous
spectrum, while a single irreducible cyclic representation of the
quantum group it has eigenvalues spaced approximately $1/\ell$ apart.
So it was proposed in \cite{Lowe:2004nw} that the appropriate quantum
representations are not the cyclic representations $\mathcal{B}$,
of dimension $N$, but rather reducible representations $\bigoplus_{i=1}^{N}\mathcal{B}_{i}$
of dimension $N^{2}$. There is one phase parameter of the cyclic
representation not fixed by matching the principal series's Casimirs,
and the sum is over different choices of this phase. In the resulting
twisted representation, $-i(L_{o}+\overline{L}_{0})$ has eigenvalues
spaced $\sim1/N\ell$, thus tending to a continuum in the classical
limit.

The natural choice for $N$ is the de Sitter radius in Planck units:
we set \[
N=\frac{\ell}{G}.\]
The maximum eigenvalue of $J_{1}$, the generator of rotations about
the south pole, is of order $N$, so this can be viewed as allowing
only those rotations which move the most distant points by at least
one Planck length. (The same $N$ would be obtained by the argument
used in \cite{Lowe:2005de}. There, the semiclassical entropy is used
to obtain an estimate of the mass gap, \cite{Preskill:1991tb} which
is then matched to the spacing of the Hamiltonian's eigenvalues.)

Note that the dimension of the twisted representation, $N^{2}=\ell^{2}/G^{2}$,
is equal to the ratio of the Planck density $1/G^{3}$ to the vacuum
energy density $\Lambda/G$. Thus the dimension of the Hilbert space
associated with a single twisted representation is essentially one
bit per unit Planck volume. In the next section, the horizon entropy
(one bit per unit Planck area) is identified as the entropy of a thermal
ensemble inside the full Hilbert space built out of tensor products
of these representations.

\section{Entropy of Kerr-de Sitter space\label{sec:Entropy-of-Kerr--de}}

Kerr-de Sitter is obtained by placing a spinning point mass at the
origin of the static patch, changing the metric to\begin{equation}
ds^{2}=-\mathcal{N}dt^{2}+\frac{dr^{2}}{\mathcal{N}}+r^{2}\left(d\phi-\frac{4GJ}{r^{2}}dt\right)^{2}\label{eq:Kerr-dS metric}\end{equation}
where\[
\mathcal{N}=M-\frac{r^{2}}{\ell^{2}}+\frac{16G^{2}J^{2}}{r^{2}}.\]
The point has mass $E=(1-M)/8G$ and angular momentum $J$. There
is still only one horizon, at radius \[
r=r_{+}=\tfrac{1}{2}\left(\sqrt{\tau}+\sqrt{\overline{\tau}}\right)\ell\]
where $\tau=M+i8GJ/\ell$. It carries entropy\begin{equation}
S=\frac{A}{4G}=\frac{\pi\ell}{4G}\left(\sqrt{\tau}+\sqrt{\overline{\tau}}\right).\label{eq:BH-Entropy}\end{equation}

This space is a quotient of pure dS$_{3}$ by a discrete group, so
is locally the same. In particular, it has the same $\Lambda$ and
the same Lie algebra of isometries. We therefore use the same quantum
deformation of this algebra, including the same $q$.

As for rotating black holes in flat space \cite{Birrell:1982ix},
the rotation creates an angular potential $\Omega$ conjugate to $J$,
in addition to the temperature $T$. The Boltzmann factor becomes
\cite{Bousso:2001mw} \[
e^{-\frac{\mathcal{M}+\Omega\mathcal{J}}{T}}=e^{\beta iL_{0}+\overline{\beta}i\overline{L}_{0}}\]
where the complex inverse temperature is given by \[
\beta=\frac{1+i\Omega}{T}=\frac{2\pi\ell}{\sqrt{\overline{\tau}}}.\]

Now consider a field living in the above twisted representation. We
propose that the microscopic CFT is formulated in terms of elementary
degrees of freedom in the twisted representation discussed above.
We will also assume that we are in a regime where free-field calculations
in the CFT suffice to give a good approximation to the entropy. In
this case, the multiparticle thermodynamic averages are as follows
\footnote{Here we assume Boltzmann statistics for simplicity, Fermi-Dirac or
Bose-Einstein changes overall numerical factors only. Note also that
for convenience the trivial tensor product is used rather than the
usual coproduct of the quantum group \cite{Klimyk:1997eb}. We have
checked via numerical calculation that this does not change the expression
for the entropy for large $N$.%
}:\begin{align*}
\left\langle -iL_{0}\right\rangle  & =\sum_{-iL_{0}>0}e^{\beta iL_{0}}(-iL_{0})\\
 & \approx N\ell\int_{0}^{\infty}dE\: e^{-\beta E}E\sim\frac{N\ell}{\beta^{2}}.\end{align*}
The states $-iL_{0}<0$ are the ones traced over to produce this thermal
behaviour. ($\sim$ here means equal up to numerical factors of order
1.) Similarly $\left\langle -i\overline{L}_{0}\right\rangle \sim N\ell/\overline{\beta}$.
Notice that unitarity \eqref{eq:so31} is restored at the level of
thermal expectation values: $\left\langle i\overline{L}_{0}\right\rangle $
is the complex conjugate of $\left\langle iL_{0}\right\rangle $,
matching $\left(iL_{0}\right)^{\star}=+i\,\overline{L}_{0}$. 

Using the relationships $\left\langle -iL_{0}\right\rangle =-\frac{\partial}{\partial\beta}\log Z$
and $\left\langle -i\overline{L}_{0}\right\rangle =-\frac{\partial}{\partial\overline{\beta}}\log Z$,
where $Z$ is the partition function, we can then calculate the entropy
as follows:\[
S=\left(1-\beta\frac{\partial}{\partial\beta}-\overline{\beta}\frac{\partial}{\partial\overline{\beta}}\right)\log Z\sim N\ell\left(\frac{1}{\beta}+\frac{1}{\,\overline{\beta}}\right)\]
matching the semiclassical result \eqref{eq:BH-Entropy} up to an
overall numerical factor %
\footnote{In AdS/CFT there is a similar discrepancy in overall numerical coefficient
in comparing the weakly coupled CFT calculation at finite temperature
with the Bekenstein-Hawking entropy of the black hole.%
}.

\section{Conclusions \label{sec:Conclusions}}

In this paper, a quantum deformed CFT has been proposed as the holographic
dual to a theory of gravity in a de Sitter background. This accounts
for the Bekenstein-Hawking horizon entropy of de Sitter spacetime.
The functional dependence on three independent parameters: cosmological
constant, mass, and angular momentum, was reproduced precisely. This
should be regarded as a very interesting success of this approach,
as other approaches to quantizing gravity in a de Sitter background
lead to divergent horizon entropy. One of the interesting features
of this construction is that the entropy is really to be thought of
as a thermal entropy, or more precisely the entropy in a canonical
ensemble with fixed temperature and angular potential. On the other
hand, the microcanonical entropy, with fixed total mass and angular
momentum, will not agree with the canonical ensemble. Instead, the
fundamental degrees of freedom are non-unitary with respect to the
standard inner product of quantum fields in de Sitter space, which
leads to imaginary angular momenta. Only when they are combined as
an ensemble with fixed angular potential does the average total angular
momentum agree with the macroscopic value of the Kerr de Sitter space.

These facts point to the instability of de Sitter spacetime. The analog
of heat baths are needed for the fixed temperature and angular potential
ensemble to make sense. The fact that we cannot ignore the presence
of these heat baths at large $N$ and obtain agreement with the microcanonical
ensemble suggests that the CFT is not a complete self-contained description
of quantum gravity in a de Sitter background. Including other degrees
of freedom becomes a necessity, which then opens the door to the complete
theory describing more than just asymptotically de Sitter spacetime.

It would be interesting to understand whether these facts relate to
the metastability of de Sitter backgrounds in string theory \cite{Kachru:2003aw}.
The hope is that the type of formulation of dS/CFT described in the
present work will provide an effective description of physics around
such backgrounds for timescales smaller than the lifetime of the de
Sitter phase.

\begin{acknowledgments}
This research is supported in part by DOE grant DE-FG02-91ER40688-Task
A.
\end{acknowledgments}
\bibliographystyle{brownphys}
\clearpage\addcontentsline{toc}{chapter}{\bibname}\bibliography{de-sitter}

\end{document}